# CHIRAL SYMMETRY BREAKING
# FROM DYSON-SCHWINGER EQUATIONS†

J.R. CUDELL

*Physics Department, McGill University, 3600 University Street*
*Montréal, Québec H3A 2T8, Canada*
cudell@hep.physics.mcgill.ca

ABSTRACT

I report on recent progress in the study of mass generation in non-abelian gauge theories, and concentrate on the origin of constituent masses in QCD. I argue that a consistent formalism has been developed to describe chiral symmetry breaking via multiplicatively renormalisable Dyson-Schwinger equations. Its main consequence is the existence of a critical value of the coupling beyond which massless solutions do not exist.

It is well-known that quarks are described by two different kinds of masses. First of all, there are the masses which enter as parameters in the lagrangian, or in sumrules, the *current masses*. For light quarks, these are small parameters ($m_u \sim 0.005$ GeV, $m_d \sim 0.01$ GeV). On the other hand, the kinematics of quarks (*e.g.* in B or D decays) and the masses of hadrons imply much larger masses for quarks, the *constituent masses*, of the order of 0.3 GeV for $u$ and $d$ quarks. These masses come from the interactions of quarks propagating in a non-trivial vacuum, similar to the interactions of electrons with the background field of a crystal. Hence 97% of the observed hadronic mass comes from the interaction of quarks with the QCD vacuum.

In the limit where one neglects current masses, the $SU(3)$ symmetry of the QCD lagrangian gets doubled into a left- and a right-$SU(3)$, as the left-handed and right-handed chiralities do not mix without a mass term. Chirality then becomes a conserved quantity, and the QCD lagrangian acquires a chiral symmetry. One then expects the *dynamics* of the theory to break this symmetry into and $SU(3)$ with mass terms, appearing as poles in the quark propagators. We shall write the quark propagator as

$$S(q^2) \equiv \frac{1}{a(q^2)\gamma \cdot q + \Sigma(q^2)} \equiv F(q^2)\gamma \cdot q + G(q^2) \qquad (1)$$

Note that we have Wick-rotated to Euclidean space, where we shall work hereafter. Chiral symmetry breaking occurs if $\Sigma(k^2)$, or equivalently $G(k^2)$, are nonzero. Note that the question we are addressing is that of mass generation in gauge theories, and the formalism we are going to develop can be applied to any gauge theory (*e.g.* to technicolour).

---

†Talk given at MRST-94 "What Next? Exploring the Future of High-Energy Physics", McGill University, Montréal, May 11-13, 1994.

Let us first explain why Dyson-Schwinger equations are relevant in this context. First of all, perturbation theory, at least in lowest order, fails to produce a mass for massless objects: the one-loop corrections to the propagator merely multiply the propagator $1/\gamma \cdot p$ by a function of $p^2$, and do not shift the position of the pole. In fact, a renormalisation group argument probably due to Callan, Dashen and Gross [1], indicates that dynamical mass generation lies in the non-perturbative regime of QCD. Indeed, the quark mass $m = \sqrt{\Sigma(m^2)}$ can only be a function of the coupling constant and of the renormalisation point $\mu$: $m(g,\mu) = \mu M(g)$, with $M$ a dimensionless function. If the mass is to be physical, it cannot depend on $\mu$, hence $\mu dm/d\mu = m + \beta(g)\partial m/\partial g = 0$, with $\beta = \partial g/\partial \mu$. In QCD with two flavours, the one-loop $\beta$- function is $\beta(g) \approx -\frac{29}{48\pi^2}g^3$, which leads to:

$$m(\mu,g) \approx \mu \exp(-\frac{24\pi^2}{29g^2}) \approx \exp(-\frac{0.6}{\alpha_S}) \qquad (2)$$

Such a dependence on $\alpha_S$ cannot be produced via perturbation theory, even if resummed, as $m$ does not have an asymptotic expansion as $\alpha_S \to 0$. Hence the problem is intrinsically non-perturbative. A caveat is here in order. It was mentioned above that perturbation theory fails to produce a mass. This is true only if one is careful about gauge invariance. Any additional term, regularization method, or resummation ansatz that breaks gauge invariance will in general produce a mass term, which will come, not from the physics, but rather from the clumsiness of the method. Hence the preservation of gauge invariance is of utmost importance in this problem.

Field theory contains few nonperturbative statements. Some of these, particularly suited for the present study, are the equations of motion of Green's functions, the Dyson-Schwinger (DS) equations. We sketch in Fig. 1 the equation for the two-point function, $S(p^2)$ which we expect to have a pole at the physical mass of the quark.

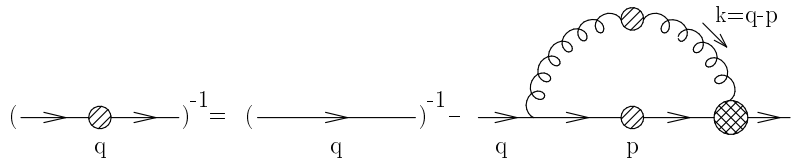

Figure 1: A pictorial representation of the Dyson-Schwinger equation for quarks. The exact two-point functions are indicated by a hatched circle, the exact three-point function by a doubly hatched circle.

Note that we work in the axial gauge, hence no ghost diagrams are present. This DS equation relates the exact quark propagator to two *a priori* unknown quantities: the exact gluon propagator and the exact three-point function. We shall assume that the gluon propagator is independently known, *e.g.* from the gluon DS equation in the

quenched approximation [2], and we shall write it as

$$D_{\mu\nu}(q^2) = q^2 D(q^2) D^0_{\mu\nu}(q^2) \tag{3}$$

with $D^0_{\mu\nu}$ the perturbative gluon propagator. We shall here rather concentrate on the approximations one can make to the three-point function. One has to guess its form, and three levels of approximation are possible. First of all, the simplest assumption is to take the perturbative vertex and assume that this is good enough. This leads to the "rainbow approximation", and to the breakdown of gauge invariance. The second level of approximation [3] is to use another non-perturbative statement, the Ward-Takahashi (WT) identities, and choose a vertex that will respect these, and hence not explicitly break gauge invariance. As we shall see, this is not restrictive enough, and one needs to go to the third level in the imposition of gauge invariance, and demand one further constraint [4], namely the matching to perturbation theory at high $q^2$, in order to obtain a sensible equation for chiral symmetry breaking.

Let us first concentrate on WT identities, as shown in Fig. 2. One can see that the projection of the three-point function onto the gluon momentum is related to the difference of two inverse propagators. Hence the part of the vertex function which is parallel to the gluon momentum, i.e. the longitudinal part, is known. The choice of vertex function is then reduced to guessing the transverse part $\Gamma_T$, which does not enter the WT identities.

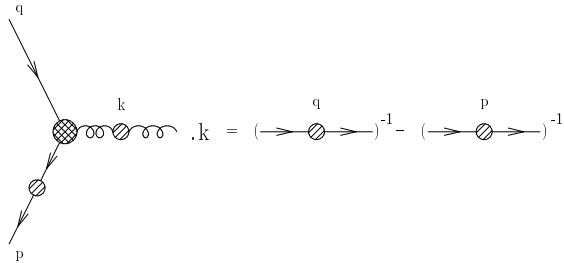

Figure 2: A pictorial representation of the Ward-Takahashi identities using the same convention as in Fig. 1.

The simplest assumption [3] is to take $\Gamma_T = 0$. One then obtains [5] two linear decoupled equations for $F$ and $G$:

$$1 = F(q^2) + \alpha_S F(q^2) \int dk^2 K_1(k^2, q^2) + \alpha_S \int dk^2 K_2(k^2, q^2) F(k^2) \tag{4}$$

$$0 = G(q^2) + \alpha_S G(q^2) \int dk^2 K_1(k^2, q^2) + \alpha_S \int dk^2 K_2(k^2, q^2) G(k^2) \tag{5}$$

The kernels $K_1$ and $K_2$ are the same in each equation, and can be found in Ref. [5]. They are both proportional to the exact gluon propagator, and hence to $D(q^2)$ of Eq. (3). Their most important property is that the integral of $K_1$ diverges logarithmically for large $k^2$: one encounters divergences similar to those of perturbation theory, which need to be renormalised away.

The renormalisation proceeds along the lines of the usual prescription: one introduces a renormalisation point $\mu$ such that $F(q^2) = Z_f(\mu^2)F_R(q^2)$, and one chooses $\mu^2 F_R(\mu^2) = 1$. As there is no quark mass term in our lagrangian, the mass squared $\Sigma$ of Eq. (1) does not get renormalised. This amounts to saying that $G$ and $F$ have the same renormalisation factor $Z_f(\mu^2)$. One also has to renormalise the gluons, which is done by similarly renormalising the gluon propagator (3): $D(q^2) = Z(\mu^2)D_R(\mu^2)$. The (infinite) renormalisation constant $Z(\mu^2)$ is absorbed in the definition of the renormalised coupling. One then subtracts from Eqs. (4,5) their value at $q^2 = \mu^2$ and this subtraction removes the remaining renormalisation constant $Z_f(\mu^2)$.

One runs however into several problems:

• In the usual multiplicative renormalisation (MR), one defines the renormalised coupling as $\alpha_S^R(\mu) = Z(\mu)\alpha_S^{bare}$, and this leads to the renormalisation group. Here however, this definition is not sufficient to remove $Z(\mu)$ from the DS equation. One has to use a definition that breaks MR:

$$\alpha_S^R(\mu) = \frac{Z(\mu)\alpha_S^{bare}}{1 - Z(\mu)\alpha_S^{bare} \int dk^2 K_1(k^2,\mu^2)} \qquad (6)$$

As the kernel $K_1$ is specific to the quark equation, one looses the universality of the renormalised QCD coupling: gluons couple differently from quarks! Furthermore, one also obtains the wrong asymptotic behaviour for $\alpha_S(q^2)$ at large $q^2$.

• The renormalised equation for $G(q^2)$ at large $q^2$ is dominated by the terms resulting from $K_1$, and one can show that these behave like $\log(\log(q^2))$. Hence the leading terms of Eq. (5) look like $G(q^2) = \alpha_S(\mu)G(q^2)\log(\log(q^2))$, and the equation has no consistent nonzero solution.

• If one proceeds and solves the equation for $F$, one obtains a solution for any $\alpha_S(\mu)$ and these solutions have a pole at the origin. Hence the quarks are massless and unconfined.

All these problems come from the large $k^2$ region, which is where we expect perturbation theory to hold. In the region where $q^2 \approx k^2 >>> p^2$, one can calculate the one-loop perturbative corrections to the vertex, and hence, by subtracting the longitudinal vertex derived from the WT identities, get an asymptotic expression for $\Gamma_T$:

$$\lim_{q>>>p} \Gamma_T^\mu(p,q) = \frac{\alpha_S \xi \log(q^2/\Lambda^2)}{4\pi q^2}(-q^\mu \gamma \cdot p + \gamma \cdot q \gamma^\mu \gamma \cdot p) \qquad (7)$$

where $\xi$ is the quark anomalous dimension. Furthermore, one has the constraint that by definition, for any $p$ and $q$, $\Gamma_T \cdot (q - p) = 0$. One can then show that the general form of $\Gamma_T$, for $\Sigma = 0$, is:

$$S^{-1}\Gamma_T^\mu S = [q^2 F(q^2) - p^2)F(p^2)]\frac{\gamma^\mu(q^2 - p^2) + (q^\mu + p^\mu)(\gamma \cdot p - \gamma \cdot q)}{\mathcal{D}} \qquad (8)$$

where $\mathcal{D}$ behaves like $q^4$ at large $q^2$, is symmetric in $p$ and $q$ and must be nonsingular. We choose in the following $\mathcal{D} = (q^2 + p^2)^2$.

We now concentrate on the chiral solution $G = 0$. The equation is similar to Eq. 4, with new kernels $\tilde{K}_1$ and $\tilde{K}_2$ replacing $K_1$ and $K_2$. These kernels are again given in Ref. [5]. Their essential property is that the divergence that used to be in the $K_1$ term is now shifted to $\tilde{K}_2$. As a consequence, the definition of the renormalised $\alpha_S$, Eq. (6), now contains a finite integral over $\tilde{K}_1$, and hence differs from the usual MR prescription by a finite calculable factor, hence the usual redefinition $\alpha_S^R(\mu) = Z(\mu)\alpha_S^{bare}$ now works and produces an equation free of singularities.

The renormalised equation looks as follows:

$$\phi(q^2)q^2 F_R(q^2) = \phi(\mu^2) + \int dk^2 [\tilde{K}_2(k^2, q^2) - \tilde{K}_2(k^2, \mu^2)] F_R(k^2) \qquad (9)$$

$$\text{with } \phi(q^2) = 1 - \alpha_S(\mu^2) \int dk^2 \tilde{K}_1(k^2, q^2)$$

This equation has two regimes. If $\phi(q^2) > 0$ for all $q^2$, then there is a solution, which means that $F \neq 0$, $G = 0$ is allowed and that chiral symmetry remains unbroken. On the other hand, if $\phi(q^2) = 0$ for some value of $q^2$ then there is no solution for $F$ and $G = 0$ is not allowed.

To study the positivity of $\phi$ for a general gluon propagator, one can use a Källen-Lehmann representation for the gluon propagator, $D(k^2) = \int d\sigma \rho(\sigma)/(k^2 + \sigma)$. The fact that the kernel $\tilde{K}_1$ is linear in $D(k^2)$, $\tilde{K}_1(q^2, k^2) \sim \tilde{\kappa}_1(k^2, q^2) D(k^2)$, enables us to rewrite $\phi$ as:

$$\phi(q^2) = \int d\sigma \rho(\sigma) H(\sigma, q^2) \qquad (10)$$

$$\text{with : } H(\sigma, q^2) = \frac{1}{\mu^2 + \sigma} - \alpha_S(\mu^2) \int dk^2 \frac{\tilde{\kappa}_1(k^2, q^2)}{k^2 + \sigma}$$

$H(\sigma, q^2) > 0$ implies that one has a massless solution. We show in Fig. 3 the region in $(\alpha_S, \sigma)$ space for which the function $H$ is positive definite, and the upper curve shows the values of $\alpha_S(\mu^2)$ for which $H(\sigma, \mu^2) = 0$, i.e. for which the effective coupling (6) becomes infinite, and where one expects the equation not to have a solution.

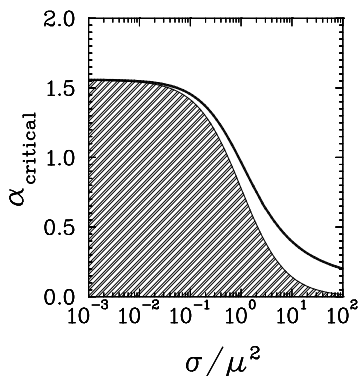

Figure 3: The shaded region is where the function $H$ is positive definite for all $q^2$; the thick curve shows the values of $\alpha_S$ for which $H(\sigma, \mu^2) = 0$.

Hence we see that massless solutions exist only for small values of $\alpha_S(\mu)$, and that there is a critical value of $\alpha_S(\mu)$ (one should really speak of a renormalisation-group invariant such as $\alpha_S(\mu)\mu^2 D_R(\mu^2)$) beyond which one ceases to have massless solutions. Futhermore, the chiral symmetry of QCD can also be destroyed by the presence of massive, high-$\sigma$ modes in the gluon propagator. Although the exact value of $\alpha_S^{critical}$ will depend on the gluon Källen-Lehmann density $\rho(\sigma)$, it is clear that a critical value of the coupling will always exist.

To show what happens to the massless solution in an explicit case, we can as a toy-model use the gluon propagator derived in Ref. [2]:

$$\mu^2 D_R(k^2) = \frac{1}{0.88 \left(\frac{k^2}{\mu^2}\right)^{0.22} - 0.95 \left(\frac{k^2}{\mu^2}\right)^{0.86} + 0.59 \, \log\left[2.1 \left(\frac{k^2}{\mu^2}\right) + 4.1\right]} \qquad (11)$$

The equation leading to this solution followed the old BBZ approach [3] and hence violated MR, and did not give rise to the renormalisation group. As a result, we were able to derive a solution only for a given $\alpha_S(\mu)$, $\alpha_S(\mu) = 1.4$, as the renormalisation group which compensates a variation of $\alpha_S$ by a variation of $D(k^2)$ did not work. The solution has a cut behaviour: near $k^2 = 0$, $D(k^2) \sim (k^2)^{-0.22}$, and hence corresponds to confined gluons, with no pole near the origin. As one expects the pole to be tapered by nonperturbative effects, and as a theorem [3] requires the axial gauge propagator to be infinite at the origin, one expects an improved solution satisfying MR to look roughly the same. Hence we use the solution found in [2] and the best we can do at present is to change the value of $\alpha_S(\mu)$ only in the quark equation. One then obtains the curves of Fig. 4. They show that as the value of $\alpha_S(\mu)$ increases, the quark propagator experiences oscillations, until eventually no solution can be found, for a critical value of $\alpha_S$ of the order of 1.5. It is interesting that the value of $\alpha_S$ that we obtained from the gluon equation is very close to the critical value.

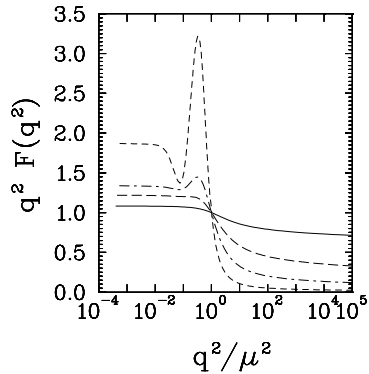

Figure 4: The solution for massless quark propagators that come from the nonperturbative gluon propagator of Ref.1, for $\alpha_S(\mu) = 0.2$ (plain), 0.6 (dashed), 1.0 (dot-dashed) and 1.4 (dashed).

To conclude, we now have a formalism to handle the Dyson-Schwinger equations which takes into account everything we know about QCD:
• The renormalisation is multiplicative, and hence gives rise to the renormalisation group equations;
• Gauge invariance is enforced via the WT identities, and via the imposition of a

transverse vertex that matches perturbative QCD in the ultraviolet region;
•The quark propagator matches smoothly with the perturbative ansatz at large momentum.

This formalism predicts that one keeps massless solutions only for small $\alpha_S$, and hence there is a critical value of $\alpha_S$ beyond which a mass must be generated for the equation to have a solution. One can also show that the equation for the mass term has a consistent behaviour at large momentum.

One of course now needs to find the solutions in the massive phase and to extend the present formalism to the gluon case. One will then have a consistent field theoretic description of chiral symmetry breaking at the quark level.

## Acknowledgements

This work was done in collaboration with A.J. Gentles and D.A. Ross and was supported in part by by NSERC (Canada), les fonds FCAR (Québec) and PPARC (United Kingdom).